\documentclass{svproc}
\usepackage{url}

\usepackage{amsmath}
\usepackage{graphicx}
\usepackage{subcaption}

\begin{document}
\mainmatter              
\title{Exploring the gluon Sivers function in photon + jet production}
\titlerunning{Exploring gluon Sivers function in photon + jet production}  
%

%
\author{
Deepesh Bhamre \inst{1} \and 
Aman Gupta \inst{2} \and 
Anuradha Misra \inst{3} \and
Siddhesh Padval \inst{3} 
}

\authorrunning{Deepesh Bhamre et al.} 
%
%

\institute{
Laborat\'{o}rio de F\'{i}sica Te\'{o}rica e Computacional-LFTC,
Universidade Cruzeiro do Sul and Universidade Cidade de S\~{a}o Paulo (UNICID),
Rua Galv\~{a}o Bueno, 01506-000, S\~{a}o Paulo, Brasil
\and
School of Physical Sciences, 
National Institute of Science Education and Research, 
An OCC of Homi Bhabha National Institute, Jatni 752050, India
\and
Centre for Excellence in Basic Sciences (UMDAE-CEBS), 
University of Mumbai, Santacruz (East), Mumbai-400098, India
}

\maketitle

\begin{abstract}
We study transverse single spin asymmetries (TSSAs) in back-to-back photon plus jet production in the scattering of unpolarized beams of protons off a transversely polarized proton target as probes of Gluon Sivers Function (GSF).
We provide estimates within the region where the imbalance $\Vec{q}_{\perp} \equiv \Vec{p}_{\gamma\perp} + \Vec{p}_{J\perp}$ between the transverse momenta of the photon and the jet is much smaller than the average of the transverse momenta
$\Vec{p}_{\perp} \equiv \lvert \Vec{p}_{\gamma\perp} - \Vec{p}_{J\perp} \rvert / 2$ 
i.e. the back-to-back region.
The presence of these two scales in the process makes it particularly suitable for applying transverse momentum dependent (TMD) factorization, in contrast to single-scale processes that rely on the generalized parton model approach.
We present our preliminary results of the transverse single spin asymmetry for this process at $\sqrt{s} $ = 200 GeV. 
\end{abstract}

\section{Introduction}
Transverse single spin asymmetries (TSSAs) observed in high energy proton-proton and lepton-proton scattering with transversely polarized protons provide valuable insights into the three-dimensional structure of the proton. 
These asymmetries are driven by the Sivers function, which is the probability of finding an unpolarized parton inside a transversely polarized proton.
Over the past years, measurements of TSSAs in semi-inclusive deep-inelastic scattering (SIDIS) from HERMES, COMPASS and JLab experiments have led to a considerate exploration of the quark Sivers function, but the gluon Sivers function is not yet well-explored.
This necessitates the need for studying processes where the gluon Sivers function can be probed. 
Photon and heavy quarkonium production processes serve as the two main processes for this purpose.

In our previous study, we considered the process
\( p^{\uparrow} p \rightarrow \gamma + X \) in Ref.~\cite{Godbole:2018mmh} 
for probing GSF.
In proton-proton collisions, prompt photons can be produced either through a hard scattering of partons or through the fragmentation of a final state parton into a photon. 
The former are referred to as {\it direct} photons, while the latter are referred to as {\it fragmentation} photons. 
At leading order $\mathcal{O}(\alpha_s\alpha_\text{em})$, direct photons are produced via the QCD Compton process \( gq \to \gamma q \), and quark-antiquark annihilation into a photon and a gluon \( q \bar{q} \to \gamma g \).
Here we consider only the direct photon contribution.
In the past, asymmetries in 
\( p^{\uparrow} p \rightarrow \gamma + \text{Jet} + X \) 
have been calculated in Refs.~\cite{Vogelsang:2005cs,Bacchetta:2007sz,Buffing:2018ggv}.
In this work, we particularly focus on the regions where the gluon Sivers function could be probed.
\section{Formalism}
We employ TMD factorization formalism for this process and write the unpolarized differential cross section as,
\begin{equation}\label{eq:unpol}
\begin{aligned}
    d\sigma&^{pp \rightarrow \gamma+\text{Jet}+X}
    =
    \frac{1}{2s}
    \sum_{ab}
    \int
    dx_{a}d^{2}\mathbf{k}_{\perp a}
    dx_{b}d^{2}\mathbf{k}_{\perp b}
    \hat{f}_{a/p}\left(x_{a}, \mathbf{k}_{\perp a}\right)
    \hat{f}_{b/p}\left(x_{b}, \mathbf{k}_{\perp b}\right)
    \\
    &
    \times
    \sum
    \frac{\lvert M\left(ab\rightarrow \gamma d\right) \rvert^{2}}
    {x_{a}x_{b}}
    \left(2\pi\right)^{4}
    \delta^{4}\left(p_{a}+p_{b}-p_{\gamma}-p_{d}\right)
    \frac{d^{3}\mathbf{p}_{\gamma}}{\left(2\pi\right)^{3}2E_{\gamma}}
    \frac{d^{3}\mathbf{p}_{d}}{\left(2\pi\right)^{3}2E_{d}}
\end{aligned}
\end{equation}
where, 
$\hat{f}_{i/p}\left(x_{i}, \mathbf{k}_{\perp i}\right)$
are the TMD parton distribution functions.

We consider transverse single spin asymmetry given by,
\begin{equation}\label{eq:AN}
    A_{N}
    \equiv
    \frac{
    d\sigma^{\uparrow}-d\sigma^{\downarrow}}
    {d\sigma^{\uparrow}+d\sigma^{\downarrow}}
\end{equation}
where $d\sigma^{\uparrow\left(\downarrow\right)}$
indicates the cross section where the proton going along the positive z axis is transversely polarized in upward (downward) direction w.r.t the X-Z plane.  
The denominator of the asymetry given by Eq.~\ref{eq:AN} is twice the unpolarized cross section (Eq.~\ref{eq:unpol}), while the numerator is given by,
\begin{equation}
\begin{aligned}
    d\sigma&^{p^{\uparrow}p \rightarrow \gamma+\text{Jet}+X}
    =
    \frac{1}{2s}
    \sum_{ab}
    \int
    dx_{a}d^{2}\mathbf{k}_{\perp a}
    dx_{b}d^{2}\mathbf{k}_{\perp b}
    \Delta^{N}f_{a/p^{\uparrow}}(x_{a},\mathbf{k}_{\perp a})
    \hat{f}_{b/p}\left(x_{b}, \mathbf{k}_{\perp b}\right)
    \\
    &
    \times
    \sum
    \frac{\lvert M\left(ab\rightarrow \gamma d\right) \rvert^{2}}
    {x_{a}x_{b}}
    \left(2\pi\right)^{4}
    \delta^{4}\left(p_{a}+p_{b}-p_{\gamma}-p_{d}\right)
    \frac{d^{3}\mathbf{p}_{\gamma}}{\left(2\pi\right)^{3}2E_{\gamma}}
    \frac{d^{3}\mathbf{p}_{d}}{\left(2\pi\right)^{3}2E_{d}}
\end{aligned}
\end{equation}
where $\Delta^{N}f_{a/p^{\uparrow}}(x_{a},\mathbf{k}_{\perp a})$ is the Sivers function.

\subsubsection{Parametrization of the TMD}\label{tmd-parametrization}
For the unpolarized TMDs, we adopt the commonly used form in which the collinear PDF is multiplied by a Gaussian transverse momentum dependence,
\begin{equation}\label{Eq:paroftmd}
f_{i/p}(x,k_\perp;Q)=f_{i/p}(x,Q)\frac{1}{\pi\langle k_\perp^2\rangle}e^{-k_\perp^2/\langle k_\perp^2\rangle}
\end{equation}
with $\langle k_\perp^2\rangle=0.25\text{ GeV}^2$ 

The Sivers function is generally parametrized as,
\begin{equation}
\Delta^{N}f_{i/p^{\uparrow}}(x,k_{\perp};Q)
=2\mathcal{N}_i(x)f_{i/p}(x,Q)\frac{\sqrt{2e}}{\pi}\sqrt{\frac{1-\rho}{\rho}}k_\perp \frac{e^{-k^2_\perp/\rho\langle k^2_\perp\rangle}}{\langle k^2_\perp\rangle^{3/2}}
\end{equation}
with $0<\rho<1$. Here $\mathcal{N}_i(x)$ is a function that parametrizes the $x$-dependence of the Sivers function. 

A commonly adopted functional form for $\mathcal{N}_i(x)$ that ensures that the positivity bound is satisfied for all values of $x$ is given by,
\begin{equation}
\mathcal{N}_i(x)=N_ix^{\alpha_i}(1-x)^{\beta_i}\frac{(\alpha_i+\beta_i)^{\alpha_i+\beta_i}}{\alpha_i^{\alpha_i}\beta_i^{\beta_i}}.
\end{equation}

In this work, in order to study the efficacy of the probe, we use Sivers functions with the positivity bound saturated, viz. $\mathcal{N}_i(x)=1$ and $\rho=2/3$. 
The parameter $\rho$ is set to 2/3 in order to maximize the first $k_\perp$-moment of the Sivers function~\cite{DAlesio:2010sag}.

\section{Results}

\begin{figure}[ht]
    \centering
    \begin{subfigure}{0.49\linewidth}
        \centering
        \includegraphics[width=\linewidth]{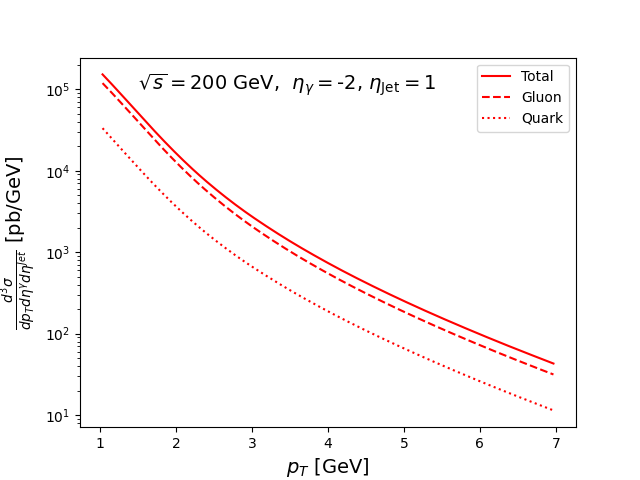}
        \captionsetup{labelformat=empty}
        \caption{Fig. 1: Unpolarized cross-section at RHIC ($\sqrt{s} = 200$ GeV).}
        \label{fig:CS}
    \end{subfigure}
    \hfill
    \begin{subfigure}{0.49\linewidth}
        \centering
        \includegraphics[width=\linewidth]{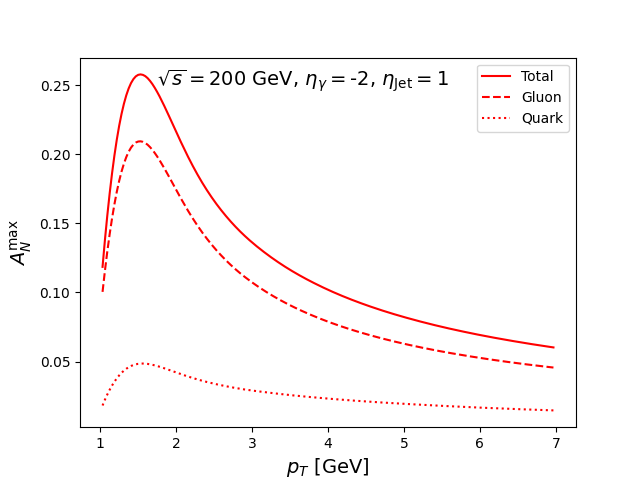}
        \captionsetup{labelformat=empty}
        \caption{Fig. 2: Estimates for Saturated asymmetry at RHIC $\sqrt{s} = 200$ GeV.}
        \label{fig:AN}
    \end{subfigure}
\end{figure}

We show the unpolarized differential cross section in Fig.~\ref{fig:CS}. 
The solid red line corresponds to the total cross section while the dashed red line and dotted red line correspond to the gluon and quark channel contribution to the total cross section respectively.
We identify the rapidity regions where the gluon channel dominates the cross section.

\vspace{-0.5cm}
We show the asymmetry results in Fig.~\ref{eq:AN}. As can be seen in the figure, the saturated form of Sivers function gives upto about 20\% asymmetry induced by the Gluon Sivers Function, and a total asymmetry upto about 25\%.

Here we mention that throughout the kinematics, we have considered 
$p_{\perp} \gg q_{\perp}$ and dropped the $\mathcal{O}\left(k_{\perp}^{2}\right)$ terms in the kinematics.
A detailed study with more precise kinematics and gauge link contribution in the Sivers function will be presented elsewhere.

\section{Conclusion}
We have identified the rapidity regions for the process 
\( p^{\uparrow} p \rightarrow \gamma + \text{Jet} + X \) 
where the GSF could be probed.
Sizeable asymmetries are found using the saturated form of the Sivers function for quark and gluon. 
The process \( p^{\uparrow} p \rightarrow \gamma + \text{Jet} + X \) is thus expected to play an important role in accessing the GSF. 
TSSAs measured for this process in the back-to-back region can be implemented for fitting the GSF. 

\section*{Acknowledgments}
D. B. acknowledges the support received from Conselho Nacional de Desenvolvimento Cient\'{i}fico e Tecnol\'{o}gico (CNPq), Brasil, Process No. 152348/2024-7. A.G. would like to thank the National Initiative on Undergraduate Science (NIUS), supported by the Department of Atomic Energy, Government of India, for financial support. A. M. and S. P. would like to thank the Department of Atomic Energy, Government of India for support through the Raja Ramanna Fellowship.


\begin{thebibliography}{6}

\bibitem{Godbole:2018mmh}
R.~M.~Godbole, A.~Kaushik, A.~Misra and S.~Padval,
Phys. Rev. D \textbf{99} (2019) no.1, 014003
doi:10.1103/PhysRevD.99.014003
[arXiv:1810.07113 [hep-ph]].

\bibitem{Vogelsang:2005cs}
W.~Vogelsang and F.~Yuan,
Phys. Rev. D \textbf{72} (2005), 054028
doi:10.1103/PhysRevD.72.054028
[arXiv:hep-ph/0507266 [hep-ph]].

\bibitem{Bacchetta:2007sz}
A.~Bacchetta, C.~Bomhof, U.~D'Alesio, P.~J.~Mulders and F.~Murgia,
Phys. Rev. Lett. \textbf{99} (2007), 212002
doi:10.1103/PhysRevLett.99.212002
[arXiv:hep-ph/0703153 [hep-ph]].

\bibitem{Buffing:2018ggv}
M.~G.~A.~Buffing, Z.~B.~Kang, K.~Lee and X.~Liu,
[arXiv:1812.07549 [hep-ph]].

\bibitem{DAlesio:2010sag}
U.~D'Alesio, F.~Murgia and C.~Pisano,
Phys. Rev. D \textbf{83} (2011), 034021
doi:10.1103/PhysRevD.83.034021
[arXiv:1011.2692 [hep-ph]].

\end{thebibliography}
\end{document}